\documentclass[twocolumn, letter]{jpsj3}
\usepackage{txfonts}
\usepackage{graphicx}
\usepackage{dcolumn}
\usepackage{bm} 
\usepackage{color}
\usepackage{ulem}

\title{Fermi Surface Deformation near Charge-Ordering Transition}

\author{Kazuyoshi Yoshimi$^{1,2}$, Takeo Kato\thanks{kato@issp.u-tokyo.ac.jp}$^{,3}$, 
and Hideaki Maebashi$^3$
}
\inst{$^1$Department of Physics, University of Tokyo, Tokyo 113-8656\\ 
$^2$Nanosystem Research Institute ``RICS'',
National Institute of Advanced Industrial Science and Technology, Ibaraki 305-8568\\
$^3$Institute for Solid State Physics, University of Tokyo, 5-1-5 Kashiwanoha, Kashiwa 277-8581
}
\abst{
We study the deformation of a Fermi surface (FS) near charge-ordering (CO) transition.
By applying a fluctuation-exchange approximation to the two-dimensional extended Hubbard model, 
we show that the FS is largely modified by strong charge fluctuations
when the wave number of the CO pattern does not match the nesting vector of
the FS in a noninteracting system. We also discuss
the enhanced anisotropy in quasiparticle properties in the resultant metallic state.
}
\kword{organic conductors, charge ordering, Fermi surface, anisotropic electron transport, 
magnetoresistance}

\begin{document}
\maketitle

The shape of a Fermi surface (FS) is an important factor in determining the electronic properties of a metal.
For many organic conductors, band structures and FSs have been obtained by
the extended H\"uckel method~\cite{Mori98,Mori99} and first-principles calculations.~\cite{Miyazaki03,Ishibashi09}
The band structures thus obtained provide a basis for constructing effective tight-binding models,
which explain the various electronic phases in different types of organic salts 
in a unified way~\cite{Seo04}. The shape of the FS is probed with high sensitivity by
transport measurements such as angle-dependent magnetoresistance
oscillations, the Shubnikov-de Haas effect and magnetoresistance~\cite{Singleton00}.

However, the shape of the FS is modified by strong electronic exchange-correlation effects, 
which are neglected in band calculations.  Deformation of the FS has already been discussed
on the basis of a single-band Hubbard model to study high-$T_{\rm c}$ superconductors. 
Theoretical studies using the second-order perturbation theory~\cite{Halboth97,Zlatic96,Noziri99},
the one-loop approximation~\cite{Yanase99} and the fluctuation-exchange (FLEX) 
approximation~\cite{Kontani99,Morita03} indicate that the FS
is gradually deformed as antiferromagnetic (AF) spin fluctuations develop near the AF transition.
In these studies, the FS is deformed so that its nesting condition improves at
the wavenumber of the AF ordering. However, the degree of deformation obtained
is small~\cite{Halboth97,Zlatic96,Noziri99,Yanase99,Kontani99,Morita03}.

In organic conductors with a $3/4$-filled band, charge fluctuations develop
near the charge-ordering (CO) transition and strongly influence the electronic properties
in a manner different from that of AF spin fluctuations~\cite{Takahashi06}. 
The effect of charge fluctuations has been clarified in recent theoretical studies of, for example, 
superconductivity mediated by charge fluctuations~\cite{Merino01, Tanaka04,Onari04, Yoshimi07}, 
non-Fermi-liquid behavior~\cite{Merino06,Cano10},
anomalous enhancement of Pauli paramagnetism~\cite{Yoshimi09} and instability toward 
inhomogeneous electronic states~\cite{Yoshimi10}. Although the electron
exchange-correlation effect is also expected to modify the FS shape
via charge fluctuations, such an effect has not yet been studied.

In this paper, we study how charge fluctuations affect the FS shape. 
By employing a FLEX approximation, we demonstrate that in contrast to
AF spin fluctuations, FS is largely modified near the CO transition. 
We emphasize that this remarkable change in FS originates from the large
discrepancy between the CO wave vector and the nesting vector
in a noninteracting system. As a result of FS deformation, 
quasiparticle properties become more anisotropic near the CO transition.
We also discuss the experimental relevance of our results.

\begin{figure}[tb]
\begin{center}
\resizebox{75mm}{!}{\includegraphics{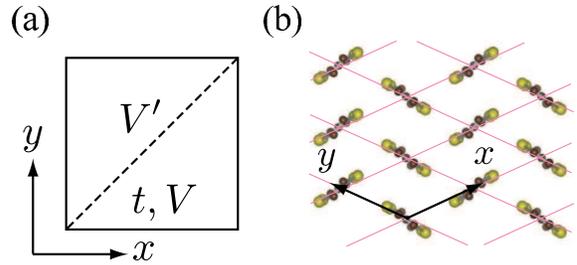}}
\end{center}
\vspace{-3mm}
\caption{(a) Schematic of the extended Hubbard model with hopping integral 
$t$ and intersite Coulomb interactions $V$ and $V'$ considered in this paper and (b) its relation to 
the crystal structure of $\theta$-ET salts.}
\label{fig:Model}
\end{figure}

We consider the extended Hubbard model on a two-dimensional square lattice. The Hamiltonian is
\begin{align}
{\cal H} \!&=\sum_{\langle i,j \rangle, \sigma} t (c_{i\sigma}^{\dagger} c_{j\sigma} \!+\!  {\rm H.c.}) 
\nonumber \\
&+\! U \sum_{i} n_{i\uparrow} n_{i\downarrow} \!
+\!  V \sum_{\langle i,j \rangle} n_i n_j \!
+\! V' \sum_{\langle \langle i,j \rangle \rangle} n_i n_j,
\label{eq:ham}
\end{align}
where $c_{i \sigma}^{\dag}$ ($c_{i \sigma}$) is the creation (annihilation) operator of 
an electron on site $i$ with spin $\sigma$ $=$ $\uparrow$ or $\downarrow$,
$n_{i\sigma}=c_{i\sigma}^{\dag}c_{i\sigma}$ and $n_{i}=n_{i\uparrow} + n_{i\downarrow}$.
Here, $t$ is the hopping integral between neighbouring sites, and
$U$ and $V$ are the onsite and nearest-neighbour Coulomb 
interactions (denoted with $\langle i,j \rangle$), respectively.
In the last term of Eq.~(\ref{eq:ham}), we have added 
an intersite Coulomb repulsion $V'$ between next-nearest-neighbour sites
in one diagonal direction, as shown in Fig.~\ref{fig:Model}~(a).

The present Hamiltonian (\ref{eq:ham}) is an effective model for CO in
the series $\theta$-ET$_2 {\rm MM}'$(SCN)$_4$ [ET=BEDT-TTF, 
M=Tl, Rb, ${\rm M}'$=Co, Zn] (abbreviated as $\theta$-ET salts)~\cite{Mori98b,Miyagawa00},
which exhibits the quasi-two-dimensional molecular arrangement shown in Fig.~\ref{fig:Model}~(b).
Owing to the crystal structure of $\theta$-ET salts, the overlap between molecular orbits is
approximately restricted to the $x$ and $y$ directions, whereas strong intersite Coulomb repulsion 
remains along one diagonal direction. The Hamiltonian (\ref{eq:ham}) has been adopted
as a minimum model in several theoretical studies of the CO phenomena in 
$\theta$-ET salts~\cite{Seo06,Kuroki09,Mori03,Watanabe06,Hotta06,Kuroki06,Nishimoto08}.

We introduce here a FLEX approximation~\cite{Bickers89} 
to study the FS deformation caused by intersite Coulomb interactions.
The one-particle Green's function $G(k)$ is related to the self-energy $\Sigma(k)$
through the Dyson equation
\begin{equation}
G(k) = \frac{1}{i \omega_n + \tilde{\mu} -\varepsilon_{\bm k} - \Sigma(k)},
\label{eq:Dyson}
\end{equation}
with a combined notation $k$ of the wave number ${\bm k}$ and the fermionic Matsubara frequency ${\rm i}\omega_n$. 
Here the noninteracting band dispersion is given by $\varepsilon_{\bm k} = 2t (\cos k_x + \cos k_y)$ and 
the Hartree term is absorbed in the chemical potential as $\tilde{\mu} = \mu-(U/2+4V+2V')n$ with the average
number $n$ of electrons per site. In the FLEX approximation, the self-energy is expressed as
\begin{align}
\Sigma(k) &= -\frac{1}{2}\int_q G(k-q) \left[ V_{c}(q)+ V_{s}(q)\right],
\label{eq:fluctuation} \\
V_{c,s}(q) &= \frac{v_{c,s}({\bm q})}{ 1+v_{c,s}({\bm q})\chi(q)},  
\label{eq:eff_Sig}
\end{align}
where $V_{c,s}(q)$ represents the exchange-correlation interaction potential,
$v_c ({\bm q})= U+4V(\cos q_x + \cos q_y)+4V'\cos (q_x +q_y)$, 
$v_s ({\bm q})=-U$, $\chi(q)= -\int_{k} G(k+q)G(k)$ and
$\int_{q} = T\sum_{\omega_n} \int_{-\pi}^{\pi} \! \int_{-\pi}^{\pi} {\rm d} {\bm q}/(2 \pi)^2$
give the combination of the sum and integral with respect to the bosonic Matsubara frequency
and momentum. We omitted the particle-hole and particle-particle ladder 
diagrams, because SU(2) symmetry of the spin sector of electrons is not relevant
if the system is far from the AF instability.\cite{Tanaka04,Onari04}
This simple approximation is sufficient to study nontrivial exchange-correlation effects on the FS.

Throughout this paper, we fix the hopping integral, the onsite interaction, and the temperature 
as $t=1$, $U=5$, and $k_{\rm B}T = 0.1$, respectively, and choose $V' = V$ for simplicity.
We checked that no AF instability occurs in the entire parameter region considered.
By combining Eqs.~(\ref{eq:Dyson})-(\ref{eq:eff_Sig}), we self-consistently 
determine $G(k)$ at 3/4 filling ($n=3/2$) with a relative precision of $10^{-6}$. 
We adopt the units $e=\hbar=1$.

\begin{figure}[tb]
\begin{center}
\resizebox{70mm}{!}{\includegraphics{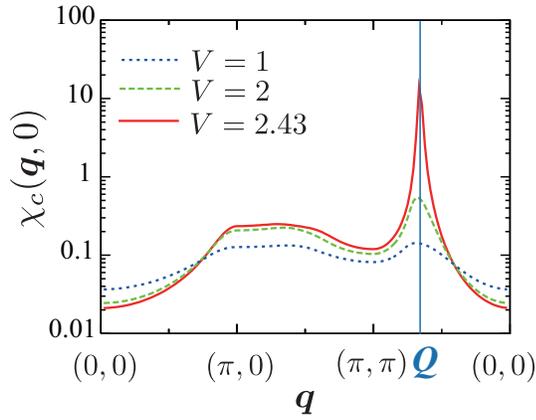}}
\end{center}
\vspace{-3mm}
\caption{Static charge susceptibility $\chi_{c}({\bm q}, 0)$ for $U=5$ and 
$k_{\rm B}T = 0.1$. The CO transition occurs at $V=2.523$. 
${\bm Q}=(2\pi/3, 2\pi/3)$ is the wave vector characterizing the CO pattern.}
\label{fig:kaic}
\end{figure}

We begin by showing in Fig.~\ref{fig:kaic} the static charge susceptibility
\begin{equation}
\chi_c({\bm q},0) = \frac{\chi({\bm q}, 0)}{1 + v_c({\bm q}) \chi({\bm q}, 0)}.
\end{equation}
As $V (=V')$ increases, the charge susceptibility increases at a specific wave vector
${\bm Q} = (2\pi/3, 2\pi/3)$, and diverges at $V = V_c = 2.523$. 
The divergence of $\chi_c ({\bm Q}, 0)$ indicates a phase transition
into three-fold-type CO~\cite{Mori03}. In the following discussion, we focus on the region 
$V \lesssim V_c$, in which charge fluctuations become large enough to modify the shape of the FS.

\begin{figure}[tb]
\begin{center}
\resizebox{70mm}{!}{\includegraphics{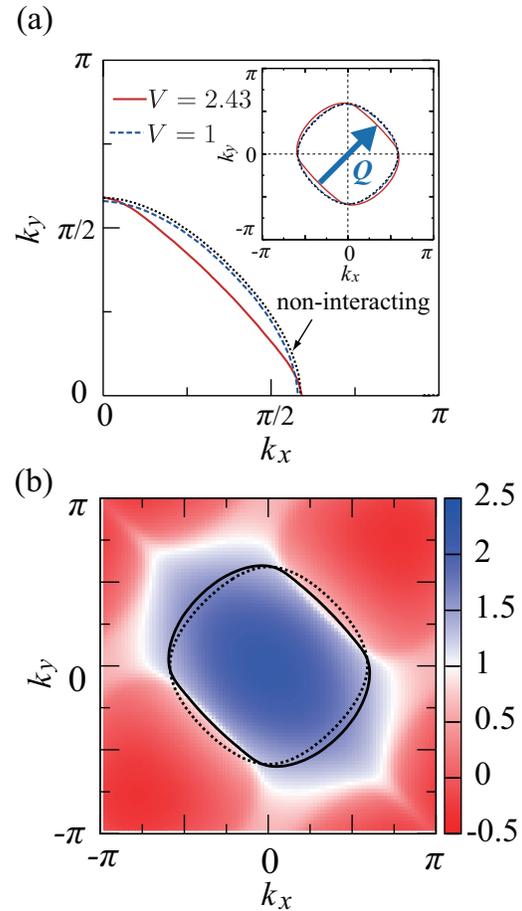}}
\end{center}
\vspace{-3mm}
\caption{(a) $V$ dependence of FS. One quarter
of the FS is plotted in the main panel with the entire FS shown in the inset. (b) 
Contour plot of ${\rm Re} \, \Sigma^R ({\bm k}, 0)$ for $V  =2.43$. }
\label{fig:FS}
\end{figure}

In Fig.~\ref{fig:FS}~(a), we show a FS determined by
\begin{equation}
\varepsilon_{{\bm k}_F} + {\rm Re} \ \Sigma^R({\bm k}_F,0) - \tilde{\mu} =0,
\label{eq:FScondition}
\end{equation}
where ${\bm k}_F$ is the Fermi wave vector and
$\Sigma^R ({\bm k}_F, 0)$ is obtained from the analytic continuation 
of $\Sigma({\bm k}_F, i \omega_k)$ in the upper plane ($\omega_k > 0$) 
by using the Pad$\acute{\rm e}$ approximant.
For weak intersite Coulomb repulsion ($V=1$), the FS (dashed line) is nearly circular, and is 
almost the same shape as that of the noninteracting electron system (dotted line).
As $V(=V')$ increases, the FS shape gradually changes.
Near the CO transition ($V=2.43$), the FS (solid line) shape clearly departs from that of the noninteracting system.
The wave vector ${\bm Q}$ of the CO pattern is also shown by the thick arrow in the inset of Fig.~\ref{fig:FS}~(a). 
The FS shape is clearly modified to {\it assist} the nesting condition; 
a flat part of the FS, which is spanned with the wave vector ${\bm Q}$, 
grows as the system approaches the CO transition. This is the main result of this study.

In the present study, the CO wave vector ${\bm Q}$ is a large mismatch for the nesting vector, which is determined 
by the shape of the FS for a noninteracting system, because ${\bm Q}$ is determined not only 
by $\chi({\rm q})$ but also by $v_c ({\bm q})= U+4V(\cos q_x + \cos q_y)+4V'\cos (q_x +q_y)$. 
However, because ${\bm Q}_{\rm AF}$ is not affected by the bare spin interaction $v_s({\bm q}) = - U$
and is determined only by $\chi({\bm q},0)$ [Eq.~(\ref{eq:eff_Sig})],
the wave vector ${\bm Q}_{\rm AF}$ of the AF fluctuations 
already properly spans the FS of the noninteracting system.
As a result, the modification of the FS induced by AF spin fluctuations is 
small compared with that induced by charge fluctuations~\cite{Halboth97,Zlatic96,Noziri99,Yanase99,Kontani99,Morita03}.

Note that the FS is already modified within the Hartree-Fock (HF) approximation, 
in which the self-energy does not include any fluctuation and is taken as
$\Sigma({\bm k}) = (-1/2)\int_{\bm q} G({\bm k}-{\bm q})(v_c({\bm q}) + v_s({\bm q}))$.
In the HF approximation, the FS is modified simply by changing the effective hopping integrals
$\tilde{t}_{ij} = t_{ij} - \sum_{\sigma} V_{ij} \langle c_{i\sigma}^{\dagger} c_{j\sigma} + {\rm h.c.} \rangle$,
where $t_{ij}$ and $V_{ij}$ are the hopping integral and Coulomb interaction 
between sites $i$ and $j$, respectively. Thus, in the HF approximation
the FS retains its round shape, whereas the FS in the FLEX approximation nests better
because of its flat part, which appears only when fluctuations are fed back into $\Sigma (k)$.

To gain a deeper understanding of how the FS deformation depends on charge fluctuations,
we expand Eq.~(\ref{eq:FScondition}) with $\tilde{\mu} = \mu_0 + \delta \mu$ and
${\bm k}_{\rm F} = {\bm k}_{{\rm F},0} + \delta k_{\rm F} \, \hat{\bm n}_{\rm F,0}$, 
where $\mu_0$ and ${\bm k}_{{\rm F},0}$ are the chemical potential and the Fermi wave vector
for the noninteracting system, respectively, and
$\hat{\bm n}_{{\rm F},0} = {\bm k}_{{\rm F},0}/|{\bm k}_{{\rm F},0}|$ is a unit vector
specifying a position on the FS~\cite{Halboth97}.
A shift in the chemical potential due to Coulomb repulsion is then given by
$\delta \mu \simeq \int {\rm d}^2 \! {\bm k} \, \delta(\varepsilon_{\rm k} - \mu_0) {\rm Re}\, \Sigma^R({\bm k},0)/\! \int {\rm d}^2 \! {\bm k} \, \delta(\varepsilon_{\rm k}- \mu_0)$, which corresponds to
an average of ${\rm Re} \, \Sigma^R({\bm k},0)$ on the noninteracting FS. 
Similarly, a shift of FS is given by
\begin{equation}
\delta k_{\rm F} \simeq \frac{\delta \mu - {\rm Re}\, \Sigma^R({\bm k}_{{\rm F},0},0)}{-|{\bm v}_{{\rm F},0}|},
\label{eq:FSdeformation}
\end{equation}
where ${\bm v}_{{\rm F},0} = \partial \varepsilon_{\bm k}/\partial {\bm k} |_{{\bm k} = {\bm k}_{{\rm F},0}}$ 
is the Fermi velocity. Here we used the fact that the direction of ${\bm v}_{{\rm F},0}$ is always
directed opposite to ${\bm k}_{{\rm F},0}$ for the present hole-type FS. 
From Eq.~(\ref{eq:FSdeformation}), the deformation direction of FS
(i.e., the sign of $\delta k_{\rm F}$) is determined by the relative magnitude of 
${\rm Re} \, \Sigma^R({\bm k},0)$ on the FS of the noninteracting system (${\bm k} = {\bm k}_{{\rm F},0}$).
In Fig.~\ref{fig:FS}~(b), we show a contour plot of ${\rm Re} \, \Sigma^R({\bm k},0)$ for $V=2.43$,
where the FSs for the interacting and noninteracting systems are shown by
the solid line and dashed line, respectively.
The results show that the deformation of the FS is mainly induced by the relative increase of the
exchange-correlation energy to holes (i.e., the decrease in ${\rm Re} \, \Sigma^R({\bm k},0)$ in terms of electrons) 
around ${\bm k} = (\pi/3, \pi/3) = {\bm Q}/2$,
which corresponds to a hot spot suffering strong electron scattering because of charge fluctuations.

\begin{figure}[tb]
\begin{center}
\resizebox{80mm}{!}{\includegraphics{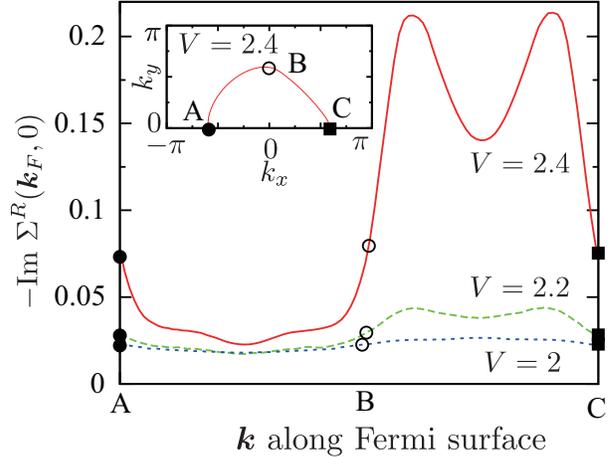}}
\end{center}
\vspace{-3mm}
\caption{Plot of $-{\rm Im} \, \Sigma^R({\bm k}_{\rm F},0)$ as a function of the position on the FS. 
Inset shows the shape of the FS for $V=2.4$ in the region of $k_y \ge 0$. 
Three representative positions on the FS are denoted by A, B, and C.}
\label{fig:sigma}
\end{figure}

As a result of FS deformation, the quasiparticle-scattering anisotropy strengthens 
because the anisotropic self-energy is enhanced near CO.
In Fig.~\ref{fig:sigma}, we show the quasiparticle scattering rate 
${\rm Im} \ \Sigma^R({\bm k}_{\rm F},0)$ as a function of ${\bm k}_{\rm F}$.
The inset of Fig.~\ref{fig:sigma} shows 
three FS points: the $-k_x$ direction (A), $+k_y$ direction (B) and $+k_x$ direction (C).
The results show that ${\rm Im} \ \Sigma^R({\bm k}_{\rm F},0)$ is largely enhanced in the {\it hot} region BC
(the flat part of the FS) as the system approaches the CO transition, whereas it remains small in the {\it cold}
region AB.  Formation of hot and cold regions reflects anisotropy 
in the exchange-correlation interaction potential, which becomes strong near the CO transition.

\begin{figure}[tb]
\begin{center}
\resizebox{80mm}{!}{\includegraphics{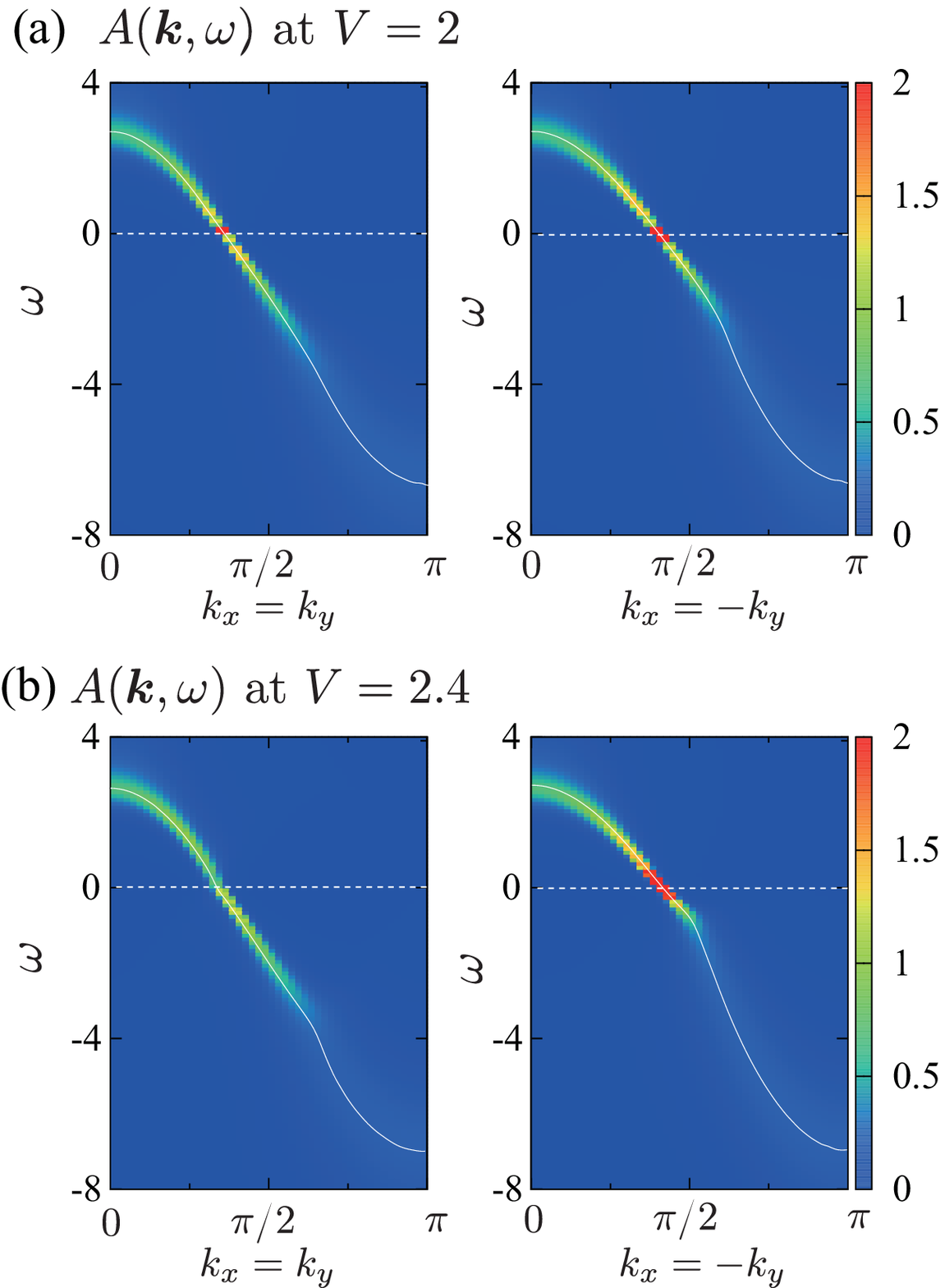}}
\end{center}
\vspace{-3mm}
\caption{Spectral weight $A({\bm k}, \omega)$ calculated for (a) $V=2$ and (b) $V=2.4$. 
The left (right) panels are contour plots of $A({\bm k},\omega)$ along the line
$k_x=k_y$ ($k_x = - k_y$). Solid white lines are the band dispersions 
determined by $\omega= \varepsilon_{\bm k} + {\rm Re} \ \Sigma^R({\bm k},\omega)$.}
\label{fig:spectrum}
\end{figure}

This anisotropic quasi-particle scattering can be observed via
the in-plane anisotropy of the electron transport such as electronic and thermal conductivities.
Another experimental probe suitable for observing both the FS deformation and anisotropic quasiparticle 
scattering is magnetoresistance. For example, out-of-plane conductivity $\sigma_{zz}$ under an in-plane 
magnetic field ${\bm B}$ may be calculated from the Boltzmann equation in the relaxation-time 
approximation:~\cite{Chambers52,Schofield00,Kovalev03,Sugawara07}
\begin{equation}
\sigma_{zz} \propto t_z^2 \int \frac{{\rm d}{\bm k}_{\rm F}}{|{\bm v}_{\rm F}|} \frac{\tau_{{\bm k}_{\rm F}}}
{1+(\Omega_{{\bm k}_{\rm F}} \tau_{{\bm k}_{\rm F}})^2},
\label{eq:magnetoresistance}
\end{equation}
where $t_z$ ($\ll \! t$) is the hopping integral between conduction planes, 
$1/\tau_{{\bm k}_{\rm F}} = 2 \, |{\rm Im} \Sigma^R({\bm k}_{\rm F},0)|$ is the
quasiparticle lifetime, ${\bm v}_{\rm F}$ is the Fermi velocity and
$\Omega_{{\bm k}_{\rm F}} = |{\bm v}_{\rm F}\times {\bm B}|/|{\bm k}_{\rm F}|$ is the cyclotron frequency.
Eq.~(\ref{eq:magnetoresistance}) indicates that
the dependence of $\sigma_{zz}$ on the orientation of the in-plane magnetic
field reflects the shape of the FS (through ${\bm v}_{\rm F}$ and ${\bm k}_{\rm F}$) as well as the anisotropy
in the quasiparticle lifetime~\cite{footnote4,footnote5}. The calculation presented in this paper thus
indicates that magnetoresistance in organic conductors is significantly affected by charge fluctuations
near the CO transition.

Finally, we discuss the spectral weight $A({\bm k},\omega) 
= - {\rm Im} \, G^R({\bm k},\omega)/\pi$, which can be measured directly by 
angle-resolved photoemission spectroscopy (ARPES) experiments.~\cite{Ito05}
We show the spectral weights for $V=2$ and $V=2.4$ in Figs.~\ref{fig:spectrum}~(a) and \ref{fig:spectrum}~(b), respectively. 
The spectral weight on the line $k_x = k_y$ (in the hot region) is largely suppressed near the Fermi energy 
when the system approaches the CO transition, while that
on the line $k_x = - k_y$ (in the cold region) near the Fermi energy remains unchanged.
Note that, below the Fermi energy, a kink structure appears on the line $k_x = -k_y$.
We attribute this high-energy kink structure to high-energy charge fluctuations~\cite{Macridin07}.

We end this paper with a brief remark on CO phenomena in $\theta$-ET salts.
In actual experiments on $\theta$-ET salts, the horizontal-type CO, which is
characterized by the wave vector ${\bm Q}_{\rm h}= (\pi/2, \pi/2)$ in our notation,
is stabilized below the CO transition temperature~\cite{Watanabe99}. The horizontal-type CO cannot be
reproduced by the RPA (or mean-field) calculation on the basis of the simple extended Hubbard model~\cite{Seo06,Kuroki09}, 
and its origin has been discussed theoretically by considering the electron-phonon interaction~\cite{Tanaka07}.
Although the horizontal-type CO cannot be obtained in our calculation, we expect that
our conclusion on the FS deformation holds true also for this type of CO, because the rapid increase of
$V_{\rm c}(q)$ toward the CO transition and the disagreement between the charge pattern and
the nesting vector are essential to our conclusion, regardless of the detailed mechanism of the CO phenomena.

In summary, we have studied the shape of the FS near the CO transition within the FLEX approximation. 
We have shown that, as the system approaches the CO transition, the FS is remarkably modified from 
that obtained from band calculations. This phenomenon is 
induced by strong charge fluctuations with a wave vector which is a large mismatch with
the nesting vector of the FS for a noninteracting system.
On the resultant FS, there appears a hot and cold region with different quasiparticle lifetimes in each region.
Our result may be observed experimentally via ARPES as well as by electron transport measurements
such as those of in-plane magnetoresistance anisotropy. Note that 
when the system is close to the CO transition, 
the FS may be sensitive to changes in temperature and/or pressure
because strength of charge fluctuations changes rapidly.
Our calculation can also be extended to more general CO models
including those of inorganic systems.

T. K. thanks H. Mori, T. Osada, S. Shin, S. Sugawara, and M. Tamura for providing useful information
on experiments with organic conductors.
This study was supported by a Grant-in-Aid for Scientific Research in Priority Area of Molecular 
Conductors (No. 21110510, 20110003, 20110004) from the Ministry of Education, Culture, Sports, 
Science and Technology, Japan.


\begin{thebibliography}{99}
\bibitem{footnote4} Note that in the previous theoretical studies, the quasiparticle lifetime
is assumed to be constant on the FS.
\bibitem{footnote5} Note that the relaxation-time approximation adopted here neglects vertex corrections,
which may change magnetoresistance quantitatively~\cite{Kontani08}. However, we
expect that neglect of the vertex corrections does not change the quantitative feature of the present result
for the anisotropy of electron transport.
\end{thebibliography}
\end{document}